\documentclass[conference, twocolumn]{IEEEtran}

\usepackage{graphicx} 
\usepackage{epsfig} 

\usepackage{steinmetz}
\usepackage{cite}
\usepackage{amsmath}
\usepackage{amssymb}
\usepackage{tikz}
\usepackage{amsthm}
\usepackage{comment}
\usepackage{caption}
\usepackage{mathcomSTEP}
\usepackage{subcaption,setspace}
\usepackage[utf8]{inputenc}
\usepackage{color}

\newcommand{\blue}[1]{\textcolor{blue}{#1}}
\newcommand{\bpcu}{\rm bpcu}

\usepackage{tikz}
\usetikzlibrary{positioning}
\allowdisplaybreaks

\newtheorem{coro}[thm]{Corollary}
\theoremstyle{remark}

\newtheorem{definition}{Definition}

 \newcommand{\I}{I}
 \newcommand{\mi}[2]{{\I}\left(#1 \, ; #2 \right)}
 \newcommand{\micnd}[3]{{\I}\left(\left. #1 ; #2 \,\right| #3\right)}

\usepackage{tikz}
\usetikzlibrary{positioning}
\usetikzlibrary{arrows,fit}

\usepackage{epstopdf}
\usepackage{pgfplots,setspace}

\tikzstyle{int}=[draw, fill=blue!10, minimum height = 1cm, minimum width=1.5cm,thick ]
\tikzstyle{int1}=[draw,  minimum height = 1cm, minimum width=1.5cm,thick ]
\tikzstyle{sum}=[circle, fill=blue!10, draw=black,line width=.5 pt,minimum size = 0.05 cm, thin ]
\tikzstyle{joint} = [draw, circle, minimum size=1em]

\newcommand{\gap}{$6.65 \ \bpcu$ }

\title{
Capacity of Discrete-Time Wiener Phase Noise Channels to Within a Constant Gap
}

\author{
	\IEEEauthorblockN{Luca Barletta}
	\IEEEauthorblockA{
		Politecnico di Milano, Italy\\
		\texttt{luca.barletta@polimi.it}
	}
	\and
	\IEEEauthorblockN{Stefano Rini}
	\IEEEauthorblockA{
		National Chiao Tung University, Taiwan\\
		\texttt{stefano@nctu.edu.tw}
	}

}

\begin{document}

\maketitle
\begin{abstract}
The capacity of the discrete-time channel affected by both additive Gaussian noise and Wiener phase noise is studied.
Novel inner and outer bounds are presented, which differ of at most $6.65$ bits per channel use for all channel parameters.
The capacity of this model can be subdivided in three regimes:
(i) for large  values of the
frequency noise variance, the channel behaves similarly to a channel with circularly uniform iid phase noise;
(ii) when the
frequency noise variance is small, 
the effects of the additive noise dominate over those of the phase  noise, while
(iii) for intermediate values of the
frequency noise variance, the transmission rate over the phase modulation channel
 has to be reduced due to the presence of phase noise.
\end{abstract}

\section{Introduction}

In the discrete-time Wiener phase noise (WPN) channel, the channel input is affected by both
 additive white Gaussian noise (AWGN) and multiplicative Wiener phase noise.
The Wiener phase process can be used to model a number of random phenomena: from imperfections in the oscillator
circuits at the transceivers, to slow fading effects in wireless environments or oscillations in the laser frequency in optical communications.
Despite its relevance in many practical scenarios, the capacity of the WPN channel remains an open problem as the presence
of memory in the phase noise process makes the analysis challenging.
In this paper we provide the first approximate characterization of capacity for the  discrete-time WPN channel
for all channel parameters and provide an input distribution which performs close to optimal.

\noindent
\emph{\bf State of the Art:} Channel models encompassing both AWGN and multiplicative phase noise have been considered in the literature
under different assumptions on the distribution of the phase noise process.
The channel model in which the phase noise is randomly selected at the beginning of transmission and is kept fixed through the transmission block-length
is  referred to as \emph{block-memoryless phase noise channel}.
In \cite{1397928} the authors prove that the capacity-achieving input distribution for this model exhibits
a circular symmetry and that the distribution of the amplitude of the input is discrete with an infinite number of mass points.
In \cite{durisi2012capacity}, the author presents capacity outer and inner bounds that capture the first two
terms of the asymptotic expansion of capacity as the signal-to-noise ratio (SNR) goes to infinity.

When the phase noise process is composed of iid circularly uniform samples, the channel is referred to as \emph{non-coherent phase noise channel}.
The capacity of this model is first studied in \cite{colavolpe2001capacity} where it is shown that the capacity-achieving
distribution is not Gaussian.
The authors of~\cite{katz2004capacity} improve on the results of \cite{colavolpe2001capacity}  by showing that the capacity-achieving
distribution, similarly to the block-memoryless phase noise channel,  is discrete and possesses an infinite number of mass points.

The WPN channel encompasses the block memoryless channel and the non-coherent channel as the two limiting cases 
in which the variance of the innovation process tends to zero and infinity, respectively.
This model was fist studied in~\cite{lapidoth2002phase} where the high SNR capacity is derived using duality arguments.
The authors of~\cite{barletta2012JLT} propose a numerical method of evaluating tight information rate bounds for this model.
In \cite{khanzadi2015capacity} the authors derive analytical approximations to capacity which are shown to be tight through numerical evaluations.

 Capacity bounds for more complex models taking into account oversampling at the receiver and/or effect of imperfect
 matched filtering are proposed in~\cite{GhozlanISIT2013,Ghozlan2014ISIT,barletta2015ISIT,barletta2015upper}:
 here the bounds are valid only at high SNR.

\noindent
\emph{\bf Contributions:} In this paper we derive the capacity of the discrete-time WPN channel to within $6.65$ bits per channel use ($\bpcu$)
for all channel parameters, namely SNR and frequency noise variance.
This result is shown by separately considering three regimes of the frequency noise variance:  \emph{small}, \emph{intermediate}, and \emph{high}  variance.
The practical insights into these regimes are as follows.
In the small frequency noise variance regime, the effects of the additive noise dominates over those of the  phase noise: for this reason the channel behaves essentially as an AWGN channel.
In the high frequency noise variance regime, the instantaneous phase variations dominate over the memory of the process
and the channel resembles  a non-coherent phase noise channel.
In the intermediate frequency noise variance regime, part of the transmission rate over the phase modulation channel has to be sacrificed
 due to the presence of the phase noise.

\noindent
\emph{\bf Organization:} The channel model and known results in literature are presented in Sec.~\ref{Sec:model}.
Some preliminary results, useful for obtaining tight capacity bounds, are detailed in Sec.~\ref{sec:Preliminary Results}.
Outer bounds are derived in Sec.~\ref{Sec:outer}, while the capacity to within a constant gap is shown in Sec.~\ref{Sec:main}.
Conclusions are drawn in Sec.~\ref{Sec:conclusion}.

\section{Channel Model and Known Results}\label{Sec:model}
The discrete-time Wiener phase noise (WPN) channel is described by the input-output relationship
\ea{
Y_i =X_i e^{j \Theta_i}+W_i , \quad i=1,\ldots,N
\label{eq:in and out}
}
where $j=\sqrt{-1}$, $W_i \sim  {\cal CN}(0,2)$, \emph{i.e.} $W_i$ is a circularly-symmetric complex Gaussian
random variable (RV) with zero mean and variance $2$. The channel input $X_i$ is subject to an average power constraint
over the transmission block-length $N$, that is
$\sum_{i=1}^N \Ebb[|X_i|^2] \leq N P$,
while $\{\Theta_i, \ i \in \Nbb\}$ is a Wiener process defined by the recursive equation
\ea{\label{eq:Wiener}
& \Theta_0=0 ,  \quad \Theta_{i+1}=\Theta_{i}+\De_{i}, \quad i\ge 0,
}
where $\De_i\sim \Ncal(0,\sgs_{\De})$, and $\sigma^2_\De$ is the
frequency noise variance.
Standard definitions of code, achievable rate and capacity are assumed in the following.

\noindent
\emph{\bf Known Results:}
As noted in \cite{lapidoth2002phase}, conditioned  on $|X_i|=|x_i|$, the modulus square of the output has a non-central chi-square
distribution with non-centrality parameter $|x_i|^2$ and two degrees of freedom, \emph{i.e.}
\ea{
|Y_i|^2  \,|\,  |X_i|=|x_i| \sim \chi_2^2(|x_i|^2).
\label{eq:chi square}
}
Moreover, in the capacity-achieving distribution, the input has iid circularly symmetric phase  $\phase{X_i},  \ i \in[1,N]$.

\begin{lem}{\bf Ergodic phase noise capacity \cite[Sec. VI]{lapidoth2002phase}.}
\label{lem:lapidoth high snr}
The capacity of the model in \eqref{eq:in and out} when $\{\Theta_i, \ i \in \Nbb \}$ is any stationary ergodic process with finite entropy rate is obtained as
\ea{
\Ccal = \f 12 \log \lb 1 + \f P 2  \rb+ \log(2 \pi)- h(\{\Theta_k\}) + o(1),
}
where $h(\{\Theta_k\}) $ is the entropy rate of the phase noise process and $o(1)$ vanishes as $P \goes \infty$.
\end{lem}
The result in Lem. \ref{lem:lapidoth high snr} establishes the high SNR capacity of the phase noise channel for a large class of phase processes
but does not apply to the WPN channel, as Wiener process is not stationary.

Specializing a  result of~\cite{barletta2015ISIT} to model~\eqref{eq:in and out} we obtain the following lemma.
\begin{lem}{\bf WPN channel capacity inner bound \cite[Sec.~III]{barletta2015ISIT}.}
	\label{lem:barletta lower}
	The capacity of the model in \eqref{eq:in and out}-\eqref{eq:Wiener} is lower-bounded as
	\begin{align}\label{eq:lowerBarletta}
		\Ccal \ge \sup_{b\ge 0, \nu>0}& \bigg\{\f 12 \log\left(\frac{P-2b+2\nu}{\pi^2 e \nu  \rho}\right) \nonumber\\
		&\quad -\left(\frac{2+b^{-1}}{\nu}+\frac{3b}{P-2b}\right)\log(e)\bigg\},
	\end{align}
	where $\rho = 1- (1-\Ebb[|X_i|^{-2}])^2 e^{-\sigma_\De^2/2}$.
\end{lem}
The inner bound in \eqref{eq:lowerBarletta} is obtained by letting the input be a truncated exponential distribution, that is
	\begin{align}
	p_{|X_i|^2}(x) = \frac{1}{P/2-b} \exp\left(-\frac{x-b}{P/2-b}\right), \qquad x\ge b.
	\end{align}


\section{Preliminary Results}
\label{sec:Preliminary Results}
In this section we present two theorems that will aid, in the following sections, the development of tight inner and outer bounds to capacity.
The first theorem bounds the entropy of a wrapped Gaussian RV while the second the entropy of a chi and chi-square RVs.

\begin{figure}
	\begin{center}
		\begin{tikzpicture}
		\node at (0,0) {\includegraphics[trim=0cm 0cm 0cm 0cm,  ,clip=true,width=\columnwidth]{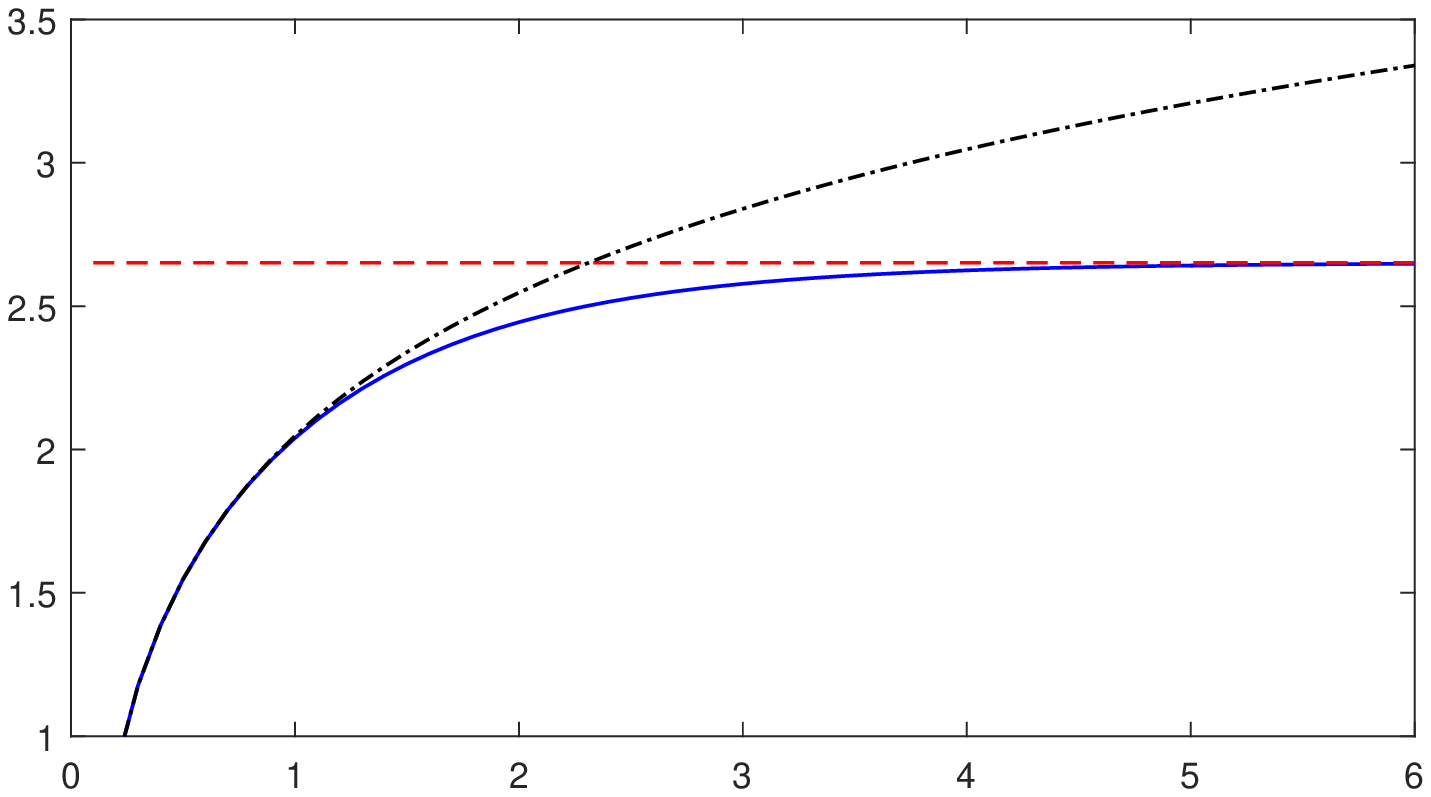}};
		\node[rotate=90] at (-4,0.3) {Entropy [bits]} ;
		\node at (0,-2.5) {$\sgs_\De$ };
		\draw[dotted] (-0.6,-2.75) node[below] {$\sgs_\De=\f {2 \pi}{e}$} -- (-0.6,+1);
		\node at (0.1,+0.2) {\blue{$h\lb  \De\rb$}};
		\node at (-2.5,+0.9) {\textcolor[rgb]{1.00,0.00,0.00}{$\log(2 \pi)$}};
		\node at (0.4,+1.5) {$\frac{1}{2}\log(2\pi e \sigma^2_\De)$};
		\end{tikzpicture}
		\caption{
The entropy of a wrapped Gaussian RV (blue solid line), a Gaussian RV (black dash-dotted line), and a uniform RV in $[0,2\pi]$ (red dashed line) as
in \eqref{eq:bound circular entropy upper} for $\sgs_{\De} \in [0,6]$.
}

		\label{fig:entropy bound}
	\end{center}
	\vspace{-.6 cm}
\end{figure}

\begin{thm}{\bf Wrapped Gaussian entropy.}
	\label{th:bound circular entropy}
	The entropy of the circularly wrapped Gaussian distribution $\De$ of variance $\sigma^2_\Delta$  is  lower-bounded as
	\ea{
		h(\De) \geq \left\{
		\begin{array}{lc}
			\log(2\pi) -2 \f {e^{-\sgs_\De}}{1-e^{-\sgs_\De}}\log(e) & \sigma^2_\Delta > 2\pi/e \\
\f 12 \log(2 \pi e \sgs_\De)+g(\sgs_\De)\log(e) & \sigma^2_\Delta \le 2\pi/e
		\end{array}
		\right.
\label{eq:bound circular entropy lower}
	}
where $g(\sgs_\De)$ is obtained as
	\begin{equation}
	g(\sgs_\De) = \f 1 2\erf\left(\frac{\pi}{\sqrt{2\sgs_\De}}\right)-\frac{e^{-\frac{\pi^2}{2\sgs_\De}}}{\sqrt{2\pi\sgs_\De}}\left(\pi+\frac{4(\pi+\frac{\sgs_\De}{\pi})}{1-e^{-\frac{\pi^2}{\sgs_\De}}}\right)-\f 1 2.
\label{eq:g bound wrapped gaussian}	\nonumber
\end{equation}
\end{thm}
\begin{IEEEproof}
	The bound involving the term $g(\sgs_\De)$ is derived in \cite{barletta2015upper}: the following derivation  is an alternative bound to
	\ea{
		h(\De) \geq \f 12 \log(2 \pi e \sgs_\De)+g(\sgs_\De)\log(e).
	}
	
	The pdf of a zero-mean wrapped Gaussian can be written, using Jacobi's triple product as
	\ea{
		p_{\De}(x)=\f 1 {2 \pi} \prod_{n=1}^{\infty} (1-q^n)(1+q^{n-1/2} e^{+i x })(1+q^{n-1/2} e^{-i x })
	}
	where $q=e^{-\sgs_\De}$ so that
	\ea{
		h(\De)  = - \log \lb \f {\phi(q)}{ 2 \pi} \rb +2 \sum_{j=1}^\infty \f {(-1)^j} {j} \f {q^{j(j+1)/2}}{1-q^j}\log(e)
		\label{eq:entropy wrapped}
	}
	where $\phi(x)$ is the Euler function.
	Note that $\phi(q)$ in \eqref{eq:entropy wrapped} is less than one by definition,
	so that $-\log(\phi(q)) \geq 0$. Moreover, the function $\kappa(j,q)$ defined as
	\ea{
		\kappa(j,q)=\f 1 {j} \f {q^{j(j+1)/2}}{1-q^j}
	}
	is decreasing in $j$ when $q \in (0,1)$ so that
	\ea{
		\sum_{j=1}^\infty \f {(-1)^j} {j} \f {q^{j(j+1)/2}}{1-q^j} \geq -\f q {1-q}.
	}

	The upper bound in  \eqref{eq:bound circular entropy upper} follows by noting that 
	\ea{
		h(\De)= h(\phase{e^{jZ}}) \le h(Z).
	}
	
\end{IEEEproof}
The bound in  \eqref{eq:bound circular entropy lower} essentially states that the entropy of the wrapped Gaussian
$\De = \phase{e^{j Z}}$ for $Z\sim \Ncal(0,\sgs_{\De})$ is well approximated as
\ea{
h\lb \De\rb \approx \min \lcb \f 12 \log(2 \pi e \sgs_\De), \log(2 \pi) \rcb,
\label{eq:bound circular entropy upper}
}
that is, the minimum between the entropy of the Gaussian RV $Z$ and a uniformly distributed RV with support $[0,2\pi]$.
As shown in Fig.~\ref{fig:entropy bound}, the approximation in \eqref{eq:bound circular entropy upper} is rather tight:
this figure plots the entropy of the  uniform RV over $[0,2\pi]$, the wrapped Gaussian RV and the Gaussian RV for different values of $\sgs_\De$.

\begin{thm}{\bf Non-central $\chi_2^2$ and $\chi_2$ entropy.}\label{th:noncentralchi2}
The entropy  of a non-central chi-square distribution with two degrees of freedom and non-central parameter $\lambda$ is bounded as
	\ea{\label{eq:bound_thm_chi2}
		\f 12  \log(8 \pi e \la)- \log(3)\leq h(\chi_2^2(\la)) \leq \f 12 \log(8\pi e(1+\lambda)).
	}
Similarly, the entropy  of a chi distribution with non-central parameter $\la$ is lower-bounded as
	\ea{
	& h(\chi_2(\la))\ge \f 12  \log(8 \pi e )-  \log(6) 
    +    \f 1 2 \Ei(-\la/2) \log(e).
\label{eq:chi square bound}
	}
\end{thm}
\begin{IEEEproof} The pdf of $T\sim \chi_2^2(\lambda)$ is 
	\ea{
		p_T(t)= \f 12 e^{-\frac{t+\lambda}{2}} I_0(\sqrt{\lambda t}) \label{eq:pT}
	}
	where 
	\eas{
		\Ebb[T]& =2+\la \\
		\var[T]& =4(1+\la).
	}

	Using the bound for $I_0(x)$ of Corollary \ref{cor:bessel}, we have
	\ea{
		h(T) & =\Ebb[-\log p_T(T)]  \\
		& \geq \Ebb \lsb 3 \log(2)+\f 1 2 \lb \sqrt{T}-\sqrt{\la}\rb^2\log(e)+ \f 14  \log(\lambda T)  \rnone \nonumber \\
		& \quad \lnone -\log(\sqrt{\pi}+1) \rsb. \label{eq:chibound}
	}
	Note that $\f 1 2 (\sqrt{T}-\sqrt{\la})^2$ is convex in $T$ for $T\ge 0$, so that we have
	\ea{
		\Ebb \lsb \f 1 2 \lb \sqrt{T}-\sqrt{\la}\rb^2 \rsb
		& \geq \f 1 2 \lb \sqrt{\Ebb[T]}-\sqrt{\la}\rb^2 \notag \\
		&= \f 1 2 \lb \sqrt{\la+2}-\sqrt{\la}\rb^2 \geq 0. \label{eq:chiboundterm1}
	}
	Furthermore, we have
	\ea{\label{eq:LogT}
		\Ebb[\log T] = \log(\la)-\Ei(-\la/2)\log(e) \geq \log(\la),
	}
	where $\Ei(\cdot)$ is the exponential integral function, 
	so that the entropy bound reads as
	\ea{
		h(T)
		& \geq \Ebb \lsb \f 12 \log(64) + \f 14  \log(\lambda^2)  \rnone \lnone -\log(\sqrt{\pi}+1) \rsb \\
		& \geq \f 12  \log\left(\frac{64}{(\sqrt{\pi}+1)^2}\lambda\right)   \notag \\
		& \geq \f 12  \log(8 \pi e \la)-\f 12 \log(3).
	}

	An upper bound on $h(T)$   can be obtained through the ``Gaussian maximizes entropy'' property as
	\ea{
		h(T)&\le \f 12 \log(2\pi e \var[T]) \notag\\
		&=\f 12 \log(8\pi e (1+\lambda)).
	}
	
	For the entropy of a chi distribution, by change of variable, we can write
	\begin{align}
	h(\sqrt{T}) &= h(T)-\Ebb[\log \sqrt{T}]-\log(2) \notag\\
	&\stackrel{\eqref{eq:bound_thm_chi2}}{\ge} \f 12  \log(8 \pi e \la)-\log(6) -\f 1 2 \Ebb [\log T] \notag\\
	&\stackrel{\eqref{eq:LogT}}{=} \f 12  \log(8 \pi e )-\log(6) +\f 1 2 \Ei(-\la/2) \log(e).
	\end{align}
\end{IEEEproof}

\begin{figure}
	\begin{center}
		\begin{tikzpicture}
		\node at (0,0) {\includegraphics[trim=0cm 0cm 0cm 0cm,  ,clip=true,width=\columnwidth]{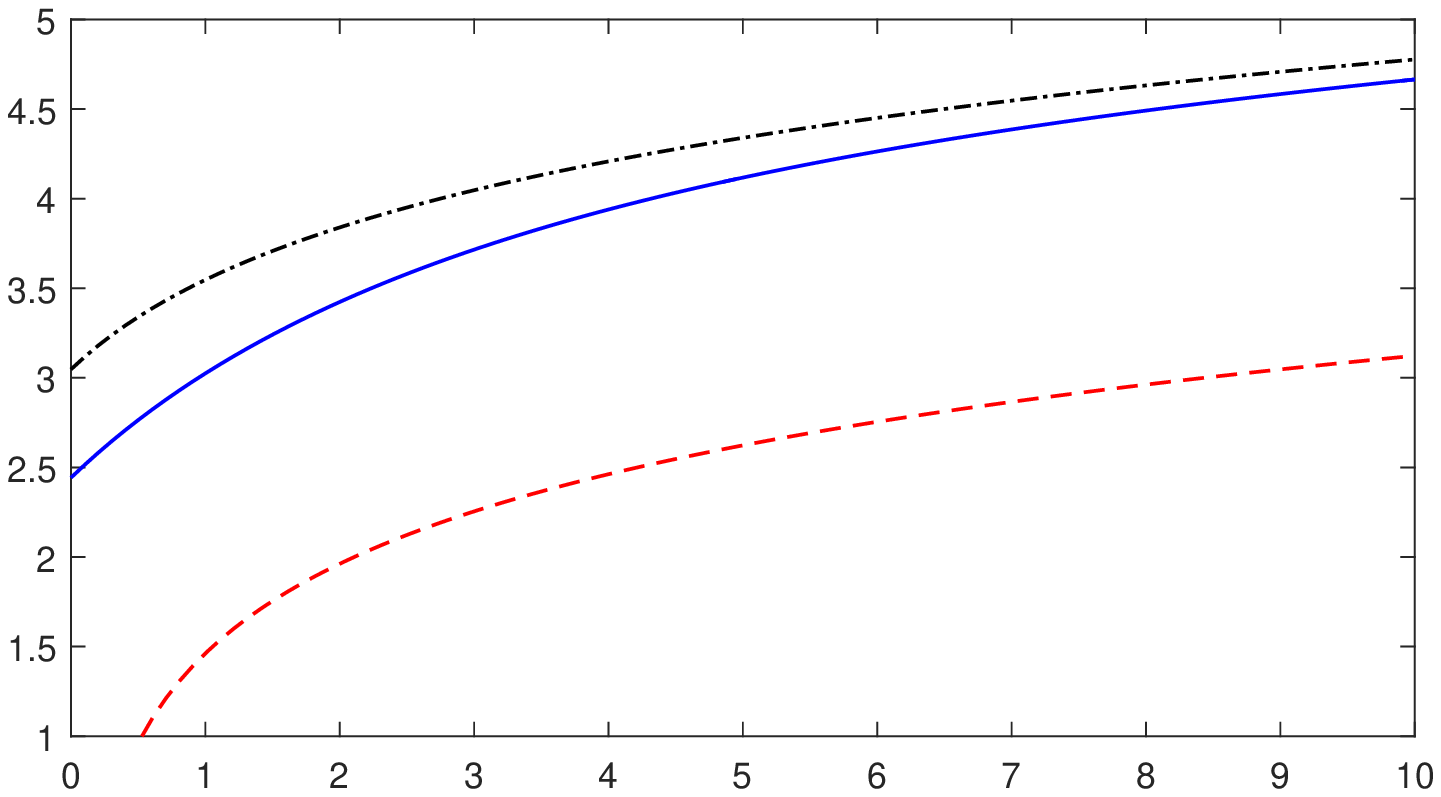}};
		\node[rotate=90] at (-4,0.3) {Entropy [bits]  } ;
		\node at (0,-2.6) {$\la$};
		\node at (0,0.5) {\blue{$h(\chi_2^2(\la))$}};
		\node at (1.5,-0.5) {\textcolor[rgb]{1.00,0.00,0.00}{inner bound}};
		\node at (-2,1.25){outer bound};
		\end{tikzpicture}
		\caption{ The entropy of a chi-square distribution (blue solid line), the inner bound (red dashed line) and the outer bound
        (black dash-dotted line) in \eqref{eq:bound_thm_chi2} for the non-centrality parameter $\la \in [0,10]$. 			
		}
		\label{fig:entropy bound chi}
	\end{center}
	\vspace{-.6 cm}
\end{figure}

The bounds~\eqref{eq:bound_thm_chi2} are plotted in Fig.~\ref{fig:entropy bound chi} as a function of the non-centrality parameter $\la$.

\section{Outer Bounds}\label{Sec:outer}
This section introduces novel outer bounds to the capacity of phase noise channel which connect this model to  the
non-coherent phase noise channel and the memoryless phase noise channel.

In the non-coherent phase noise channel, the phase noise process $\{\Theta_i, \ i \in \Nbb \}$ is an iid sequence of uniformly distributed RVs in $[0,2\pi]$:
for this reason reliable communication can take place only over the amplitude modulation channel $p_{|Y|  \,|\, |X|}$. The rate attainable over this channel
is bounded in the next theorem.

\begin{thm}{\bf Outer bound on capacity over non-coherent AWGN channel.}\label{thm:AWGNinner}
The capacity of the non-coherent phase noise channel is upper-bounded as
\ea{
\sup_{p_{X}:\,\Ebb|X|^2\le P} \mi{|X|}{|X+Z|} \le \f 1 2 \log(2\pi e(P+2)),
}
where $Z\sim {\cal CN}(0,2)$.
\end{thm}
\begin{IEEEproof}
Using the Gaussian maximum entropy bound we bound the positive entropy term as $h(|X+Z|)  \leq 1/2\log(2\pi e(P+2))$.
For the conditional entropy we write
	\begin{align}
	& h(|X+Z| \,|\, |X|)
    \ge \max_{x\ge 0} h(|X+Z| \,|\,|X|, |X|>x) \notag\\
	&\ge \max_{x\ge 0} \int_{x}^{\infty} h(|t+Z|)\: \diff F_{|X|}(t) \notag\\
	&\ge \max_{x\ge 0} \Pr(|X|>x) \left(\f 12  \log\left(\frac{8 \pi e}{6^2} \right)
	+\f 1 2 \Ei\lb-\frac{x^2}{2}\rb \log(e)\right),\nonumber
	\end{align}
	where in the last step we used the bound in Thm.~\ref{th:noncentralchi2} and the (increasing) monotonicity of the bound in \eqref{eq:chi square bound}.
	Since
	\begin{align}
	\max_{x\ge 0} f(x) &= \max \left\{ \max_{0\le x < r} f(x), \max_{x\ge r} f(x)  \right\},
	\end{align}
	and by choosing $r$ such that $\f 12  \log(8 \pi e )-\log(6) +\f 1 2 \Ei(-r^2/2) \log(e)=0$
(there exists only one value of $r$ with this property, namely $r\approx 0.937$), we conclude that $h(|X+Z| \,|\, |X|)$ must be positive,	
	since the maximization for $0\le x < r$ gives a negative number, while the maximization for $x \ge r$ gives a positive number.	
\end{IEEEproof}

The next outer bound is a refinement of the result in Lem.~\ref{lem:lapidoth high snr} to yield an outer bound to the capacity of the model
in \eqref{eq:in and out}-\eqref{eq:Wiener}
valid at finite SNRs.
The bound is derived by revealing the past phase realization to the receiver, which results in a memoryless phase noise channel.

\begin{thm}{\bf Memoryless phase noise channel outer bound.}
	\label{thm:outer bound}
	The capacity of the WPN channel in \eqref{eq:in and out}-\eqref{eq:Wiener} can be upper-bounded as
	\ea{
		\Ccal&\leq
     \min \lcb   \f 1 2 \log(2\pi e(P+2))+ \log(2 \pi) - h(\Delta),  \rnone \nonumber\\
  & \quad \quad    \lnone \log\left(1+P/2 \right) \rcb,
\label{eq:outer bound min}	
}
	where  $h(\De)$ is the entropy of a wrapped Gaussian with variance $\sigma^2_\De$.
\end{thm}
\begin{IEEEproof}
A trivial capacity outer bound is $\Ccal  \leq \log\left(1+P/2 \right)$
and is obtained by  providing the phase noise sequence to the receiver.
Another capacity outer bound is obtained as follows:
\ean{
 I(X^N;Y^N)    
&\le \sum_k I(X^N, \Theta_{k-1};Y_k|Y^{k-1})  \\
& =\sum_k \lb I(X^N;Y_k| \Theta_{k-1}) + I(\Theta_{k-1} ; Y_k | Y^{k-1}) \rb \\
& =\sum_k \lb I(X_k;Y_k| \Theta_{k-1}) + I(\Theta_{k-1} ; Y_k | Y^{k-1}) \rb .
}
Since the additive noise $W$ is circularly symmetric, a uniformly distributed input phase $\phase{ X_k}$ in $[0,2\pi)$ is capacity-achieving. This also implies that
\ean{
 &I(\Theta_{k-1};Y_k|Y^{k-1}) \\
  &= I(\Theta_{k-1};|X_k| e^{j (\Theta_{k-1}+\Delta_{k-1} + \angle{ X_k})}+W_k | Y^{k-1})  =0,
}
given that  $\phase{X_k}$ is  independent of $\Theta_{k-1}$.
	Using the polar decomposition $X=|X|e^{j \angle X}$, and dropping the time index for convenience of notation, write
	\eas{
		&\micnd{X_k}{Y_k}{\Theta_{k-1} } =  \mi{X}{X e^{j\Delta}+W}\\
		& = \mi{|X|}{|X+W|}+\micnd{\phase{ X}}{Xe^{i\Delta}+W}{|X|}
\label{eq:step 2}\\
		& \le \mi{|X|}{|X+W|}+\mi{\phase{ X}}{\phase{ X} \oplus \Delta }
\label{eq:step 3}\\
		& \le \f 1 2 \log(2\pi e(P+2))+ \log(2 \pi) - h(\Delta),
\label{eq:step 4}
	}
	where \eqref{eq:step 2} follows by circular symmetry of $W$,  \eqref{eq:step 3} by revealing $W$ to the receiver,
and \eqref{eq:step 4} from Thm.~\ref{thm:AWGNinner} and the circular symmetry of $X$.
The symbol $\oplus$ in \eqref{eq:step 3} denotes the addition modulo $2\pi$.
\end{IEEEproof}
\section{Main Result}\label{Sec:main}
\begin{thm}{\bf Capacity to within a constant gap.}
\label{th:Capacity to within a constant gap}
The capacity of the WPN channel in \eqref{eq:in and out}-\eqref{eq:Wiener} is upper-bounded as
\begin{align}
\Ccal &\leq \f 1 2 \log(1+P/2)\nonumber\\&+ \lcb  \p{
\f 1 2 \log(4\pi e)+ 2 \f {e^{-\frac{2\pi}{e}}}{1-e^{-\frac{2\pi}{e}}}\log(e) &    \sgs_\De>\f {2 \pi}{e}  \\
 \f 12 \log \lb  \f 2 {   \sgs_\De}  \rb +\log(2\pi) +\log^2(e)
&   P^{-1} \leq \sgs_\De \leq \f {2 \pi}{e} \\
\f 12 \log(1+P/2)    &  P^{-1}>\sgs_\De
} \rnone
\label{eq:outer}
\end{align}
and the exact capacity is to within $ \Gcal \ \bpcu$ from the outer bound in \eqref{eq:outer}, where
\ea{
\Gcal \leq \lcb\p{
4 &    \sgs_\De>\f {2 \pi}{e}\\
6.65 &   P^{-1} \leq \sgs_\De \leq \f {2 \pi}{e} \\
1.21 &  P^{-1}>\sgs_\De
}
\rnone
\label{eq:regimes phase noise}
}
\end{thm}

\begin{IEEEproof}
The achievability proof  relies on a simple transmission scheme employing iid complex Gaussian inputs
while the converse proof is developed from the outer bound in Thm.~\ref{thm:outer bound}.

\noindent
\emph{Achievability:}
As in~\cite[Eq.~(18)]{barletta2015ISIT}, the WPN channel capacity can be lower-bounded as
	\begin{align}
	&
	{\cal C}\ge
	 \frac{1}{N} \sum_{k=1}^{N} \micnd{|X_k|}{Y_1^N}{X_1^{k-1}} + \micnd{\phase{X_k}}{Y_1^N}{X_1^{k-1},|X_k|}
	\nonumber\\
	 &\ge \mi{|X_1|^2}{|Y_1|^2} + \micnd{\phase{X_1}}{Y_{0}^1}{X_{0},|X_1|} \label{eq:first}.
	\end{align}
	
	Next we lower-bound the term \mbox{$I_{||}=\mi{|X_1|^2}{|Y_1|^2}$} and $I_{\angle}=\micnd{\phase{X_1}}{Y_{0}^1}{X_{0},|X_1|}$, which we refer to as the amplitude and phase modulation channel, respectively.
	
\smallskip
\noindent
{$\bullet$  \bf Amplitude modulation channel:}
Let the input distribution be $X\sim {\cal CN}(0,P)$.
Recognizing that $|Y|^2 = (1+P/2) K$ where  $K \sim \chi^2_2$, we obtain
	\ea{
	\mi{|X|^2}{|Y|^2}
	&= \log\left(e(2+P)\right)  - h(|Y|^2\, | \, |X|^2).
\label{eq:amplitude 1}	
}	
Using the outer bound of Thm.~\ref{th:noncentralchi2}, we bound the term $h(|Y|^2\, | \, |X|^2)$ as
	\ea{
		h(|Y|^2\,|\,|X|^2) &\le \f 12 \Ebb \log(8 \pi e(|X|^2+1)) \le \f 1 2 \log(8\pi e (1+P)).
\label{eq:amplitude 2}
	}
Finally, combining  \eqref{eq:amplitude 1} and \eqref{eq:amplitude 2} yields
\ea{
I_{||} & \geq \f 1 2\log\left(1+\frac{P}{2}\right) - \f 1 2 \log\left(\frac{2\pi }{e}\right)+ \f 1 2 \log\left(\frac{1+P/2}{1+P}\right).
\label{eq:lower_amplitude}
}

\noindent
	{$\bullet$  \bf Phase modulation channel:}
	The second term in the RHS of~\eqref{eq:first} can be lower-bounded as follows:
	\ea{
	&I_{\angle} = \micnd{\phase{X_1}}{Y_{0}^1}{X_{0},|X_1|} \nonumber \\
	&=\micnd{\phase{X_1}}{X_0+W_0}{X_{0},|X_1|} \nonumber\\&\quad+ \micnd{\phase{X_1}}{X_1 e^{j \De_0}+W_1}{X_{0},|X_1|,X_0 + W_0} \nonumber \\
	&= \micnd{\phase{X_1}}{X_1 e^{j \De_0}+W_1}{|X_1|} \nonumber \\
	& \ge \micnd{\phase{X_1}}{\phase{X_1} \oplus \De_0 \oplus N}{|X_1|}\nonumber\\
	&=\log(2\pi) - h(\De_0 \oplus N |\, |X_1|) \label{eq:phase},
	}
	where \eqref{eq:phase} follows by considering just the phase of $Y_1$, and $N = \phase{|X_1| + W_1}$.
Since $\De_0 \oplus N$ is defined over the support $[0,2 \pi]$, we can apply the maximum entropy theorem to upper-bound the conditional entropy term
 $h(\De_0 \oplus N |\, |X_1|)$ with the entropy of a  wrapped Gaussian RV with variance $\sigma^2_\Delta + 1/|X_1|^2$:
\eas{
	&h(\De_0 \oplus N |\, |X_1|) \\
    & \leq \f 12 \Ebb\left[\log\left(2\pi e \left(\sigma^2_\Delta + \f {1} {|X_1|^2}\right)\right)\right]
    \label{eq:phase 2 1} \\
	&=\f 12 \Ebb\left[\log\left(2\pi e \left(\sigma^2_\Delta |X_1|^2 + 1\right)\right)\right]-\f 12 \Ebb\left[\log|X_1|^2\right] \nonumber\\
	&\le \f 12 \log\left(2\pi e \left(\frac{\sigma^2_\Delta P + 1}{P}\right)\right)+\f  \gamma 2 \log(e),
\label{eq:phase 2}
}{\label{eq:phase final}}
where \eqref{eq:phase 2 1} follows by Thm.~\ref{th:bound circular entropy}, and  \eqref{eq:phase 2} from Jensen's inequality for the first term
and from the fact that \mbox{$\Ebb\log|X_1|^2 = \log (P e^{-\gamma})$}, where $\gamma$ is the Euler-Mascheroni constant.

\smallskip
	
Putting together the contributions of amplitude and phase modulation in \eqref{eq:lower_amplitude} and \eqref{eq:phase final} respectively,
we obtain the inner bound
\ea{
{\cal C}
&\ge \f 1 2\log\left(1+\frac{P}{2}\right) + \f 1 2 \log\left(\frac{1+P/2}{1+P}\frac{P e^{-\gamma}}{\sigma^2_\Delta P + 1}\right).
\label{eq:inner bound}
}
	
Note that, for any fixed $\sigma^2_\Delta$, the pre-log at large $P$ given by~\eqref{eq:inner bound} is $1/2$.
Also, note that if $\sigma^2_\Delta \le 1/P$, then the pre-log at large $P$ is~$1$.

\smallskip
\noindent
\emph{Converse and gap from capacity:}
The outer bound in Th. \ref{thm:outer bound} can be sub-divided in the three regimes of small, intermediate, and high frequency noise variance.

\noindent
$\bullet$  {\bf High frequency noise variance: $\sigma_\De^2>{2 \pi}/{e}$:}
The outer bound from Th. \ref{thm:outer bound} together with the condition $\sigma_\De^2>{2 \pi}/{e}$ yields the outer bound
\begin{align}
{\cal C}&\le
\left\{
\begin{array}{lc}
\f 12\log(1+P/2)& P\le 10 \\
\f 1 2 \log(2\pi e(P+2))+ 2 \f {e^{-\frac{2\pi}{e}}}{1-e^{-\frac{2\pi}{e}}}\log(e) & P>10
\end{array}
\right.
\label{eq:outer bound High}
\end{align}
By comparing the outer bound in \eqref{eq:outer bound High} with the inner bound  in \eqref{eq:inner bound} we obtain the capacity gap
\ea{
\Gcal
& \le \left\{
\begin{array}{lc}
	\f 12\log(1+P)+ \f 1 2 \log\left(\frac{2\pi }{e}\right) & P\le 10 \\
	\f 1 2 \log\left(\frac{1+P}{2+P}\right)+ 2 \f {e^{-\frac{2\pi}{e}}}{1-e^{-\frac{2\pi}{e}}}\log(e) +  \log\left(4\pi \right) & P>10
\end{array}
\right.
}
which
 is smaller than $4 \  \bpcu$ for any $P>0$ and $\sigma_\Delta^2 >2\pi/e$.

\noindent
$\bullet$ {\bf Small frequency noise variance: $\sgs_\De<P^{-1}$:}
In this regime we consider the trivial outer bound ${\cal C} \le  \log(1+P/2)$.
When $P<2$, capacity is necessarily less than $1\:\bpcu$.
When $P\geq 2$, the gap between the trivial outer bound and the inner bound in~\eqref{eq:inner bound} is at most $1.21\, \bpcu$.

\noindent
$\bullet$ {\bf Intermediate frequency noise variance: $\frac{1}{P} \leq \sgs_\De \leq \frac{2\pi}{e} $:}
The outer bound in Th. \ref{thm:outer bound} in this regime can be rewritten as
\ea{
{\cal C} & \le \f 1 2 \log\lb\frac{P+2}{\sigma^2_\De}\rb + \log(2\pi) -g(\sgs_\De)\log(e) \nonumber\\
&\le \f 1 2 \log\lb\frac{P+2}{\sigma^2_\De}\rb + \log(2\pi)+\log^2(e)\label{eq:outer bound medium},
}
where in the last step we note that $g(\sgs_\De)$ is decreasing in $\sgs_{\De}$ so that $g(\sgs_\De) \geq -\log(e)$.
We next compare the inner bound in \eqref{eq:inner bound} with the outer bound in \eqref{eq:outer bound medium} and obtain
\begin{align}
{\cal G} 
& =\lb \f  \gamma 2+ \log(e) \rb \log(e)
+ \f 12 \log\left((2\pi)^2\frac{1+P}{2+P} \frac{1+\sigma^2_\Delta P}{\sigma^2_\Delta P}\right) \nonumber\\
& \le\lb \f  \gamma 2+ \log(e) \rb \log(e)+\log(2 \pi) + \f 3 2 \log(2) \leq 6.65, \nonumber
\end{align}
where, in the last passage, we have considered the worst case with $\sigma^2_\Delta = P^{-1}$ and $P\rightarrow\infty$.
\end{IEEEproof}

Note that the inner bound in Th. \ref{th:Capacity to within a constant gap} relies on iid complex Gaussian inputs, thus showing that this
input distribution performs sufficiently close to capacity.

Some further insights on the result in Th. \ref{th:Capacity to within a constant gap} emerge by  simplifying the outer bound in
\eqref{eq:outer} as in the following lemma.

\begin{lem}{\bf Larger gap, simpler expression.}
\label{th: Larger gap, simpler expression}
The capacity of the WPN channel in in \eqref{eq:in and out}-\eqref{eq:Wiener} is upper-bounded as
\ea{
\Ccal \leq \lcb  \p{
\f 1 2 \log(1+P/2)+ 4
&    \sgs_\De>\f {2 \pi}{e} \\
\f 12  \log(1+P/2) - \f 12 \log   \sgs_\De  +5.5
&   P^{-1} \leq \sgs_\De \leq \f {2 \pi}{e} \\
 \log(1+P/2)    &  P^{-1}>\sgs_\De
} \rnone
\label{eq:outer simpler 1}
}
and the exact capacity is to within a gap of  $7 \ \bpcu$ from the outer bound in \eqref{eq:outer simpler 1}.
\end{lem}

The three regimes in \eqref{eq:outer simpler 1} can be intuitively explained as follows.
The inequality $\sigma^2_\Delta \gtrless  2\pi/e $ in \eqref{eq:outer simpler 1} arises from the result in Th. \ref{th:bound circular entropy}
 and whether the phase noise entropy is better approximated by using a uniform RV over $[0,2\pi]$ or
a Gaussian RV.
When the frequency noise variance is high, the channel behaves similarly to a channel with uniform phase noise, in which transmission
takes place only in the  amplitude modulation channel.
When the frequency noise variance is small, instead, the effect of the multiplicative noise times the channel input has variance $P \sgs \leq 1$ 
which is smaller than the variance of the additive noise. 
The inequality $P^{-1} \leq \sgs_\De $, as in Th.~\ref{thm:outer bound}, intuitively arises from the rate attainable on the phase modulation channel:
the largest rate attainable on this channel is close to $1/2\log(P+1)$, when the  frequency noise variance is relatively small.
For a higher  frequency noise variance, the rate of the phase modulation channel is instead close to $1/2 \log( \sigma^{-2}_\Delta)$.

\begin{figure}
	\begin{center}
		\begin{tikzpicture}
		\node at (0,0) {\includegraphics[trim=0cm 0cm 0cm 0cm,  ,clip=true,width=\columnwidth]{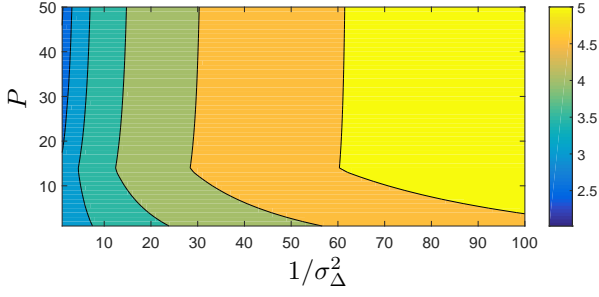}};
		\node[rotate=90] at (-4,0.3) {$P$} ;
		\node at (0,-2) {$1/\sgs_{\De}$};
		\end{tikzpicture}
		\caption{
A contour plot of the exact gap between inner and outer bound in the proof of Th. \ref{th:Capacity to within a constant gap} for $P\in[0,50]$ and  $1/\sgs_{\De}\in [1,100]$.	
		}
		\label{fig:GapPlotII.eps}
	\end{center}
	\vspace{-.6 cm}
\end{figure}

Although the largest gap between inner and outer bound is bounded by \gap, the difference between inner and outer bounds for any parameter regime
 can be easily evaluated from the proof of Th. \ref{th:Capacity to within a constant gap}.
A plot of the exact value of $\Gcal$ in~\eqref{eq:regimes phase noise} is
 presented in Fig. \ref{fig:GapPlotII.eps}.

\section{Conclusions}\label{Sec:conclusion}
We have derived outer and inner bounds of the discrete-time Wiener phase noise channel and have shown that they differ of at most \gap \ at any SNR
and frequency noise variance.
Both bounds have rather simple expressions and suggest that all channels can be roughly divided in three parameter regimes:
 high, intermediate, and small frequency noise variance.
 Moreover, the analysis of the inner bound shows that a complex Gaussian input distribution  performs rather close to capacity.

\appendix
\begin{lem}{\bf Upper bound on modified Bessel function\\}\label{th:bessel bound}
	The zero-th order modified Bessel function of the first kind can be upper-bounded as
	\ea{I_0(x) &\le
		\f {e^x} {4\sqrt{x}} \lb \erf(\sqrt{x})  \sqrt{\pi}+\frac{2e^{-x}(1-e^{-x})}{\sqrt{x}} \rb
		\label{eq:bessel bound}
	}
	for any $x\ge0$, where 
	\ea{
		\erf(x) = \f 2 {\sqrt{\pi}} \int_{0}^{x} e^{-t^2} \diff t.
	}
\end{lem}
\begin{IEEEproof}
	By definition
	\ea{
		I_0(x) &= \frac{1}{\pi} \int_{0}^{\pi} e^{x\cos(\theta)} d\theta  \notag \\
		&=\underbrace{\frac{1}{\pi} \int_{0}^{\pi/2} e^{x\cos(\theta)} d\theta}_{ {\cal I}_1} +\underbrace{\frac{1}{\pi} \int_{\pi/2}^{\pi} e^{x\cos(\theta)} d\theta}_{{\cal I}_2}.
	}
	Using the infinite product formula for the cosine function and a linear lower bound for the cosine, we obtain the bound
	\ea{
		e^{x \cos(\theta)} & \leq  \left\{
		\begin{array}{lc}
			e^{x(1-\f 4{\pi^2} \theta^2)} & 0\le \theta\le \pi/2\\
			e^{x(1-\f 2 {\pi}\theta)} & \pi/2\le \theta\le \pi
		\end{array}
		\right.
	}
	so that
	\ea{
		{\cal I}_1
		& \leq \frac{1}{\pi}\int_{0}^{\pi/2} e^{x\lb 1-\f 4{\pi^2} \theta^2 \rb } d\theta \notag\\
		& = \f {\sqrt{\pi}} 4 \f {e^x } {\sqrt{x}} \erf(\sqrt{x}) 
		\label{eq:bound I1}
	}
	while
	\ea{
		{\cal I}_2
		&  \leq \frac{1}{\pi}\int_{\pi/2}^{\pi} e^{x\lb 1-\f 2 {\pi} \theta \rb } d\theta \notag\\
		& =  \f {1-e^{-x}} {2x}.
		\label{eq:bound I2}
	}
	Combining  \eqref{eq:bound I1} and \eqref{eq:bound I2} we obtain the bound
	\ea{
		I_0(x)
		& \leq \f {e^x} {4\sqrt{x}} \lb \erf(\sqrt{x})  \sqrt{\pi}+\frac{2e^{-x}(1-e^{-x})}{\sqrt{x}} \rb
		\label{eq:bound I0 erf}
	}
	
\end{IEEEproof}

The result in Lemma~\ref{th:bessel bound} can be further weakened to obtain an expression which can be more easily manipulated analytically.

\begin{coro}\label{cor:bessel}
	An upper bound to $I_0(x)$ is
	\begin{equation}
	I_0(x)\le \frac{e^x}{\sqrt{x}} \frac{\sqrt{\pi}+1}{4}.
	\label{eq: cor bessel}
	\end{equation}
\end{coro}
\begin{IEEEproof}
	From Lemma \ref{th:bessel bound} we have
	\begin{align}
	I_0(x) &\le \f {e^x} {4\sqrt{x}} \lb \erf(\sqrt{x})  \sqrt{\pi}+\frac{2e^{-x}(1-e^{-x})}{\sqrt{x}} \rb \nonumber\\
	&\le \f {e^x} {4\sqrt{x}} \lb   \sqrt{\pi}+1 \rb
	\end{align}
	where the last equality follows by $\erf(x)\le 1$ and the fact that $2e^{-x}(1-e^{-x})< \sqrt{x}$.
\end{IEEEproof}
Note that, although simpler, the expression in \eqref{eq: cor bessel} is not tight as $x \goes 0$, where it actually has an asymptote.
The expression in  \eqref{eq:bessel bound}, although cumbersome, is tight as $x \goes 0$.

\bibliographystyle{IEEEtran}
\bibliography{steBib1}

\end{document}